# Intermediate ferroelectric orthorhombic and monoclinic $M_B$ phases in [110]- electric field cooled Pb(Mg$_{1/3}$Nb$_{2/3}$)O$_3$-30%PbTiO$_3$ crystals

Hu Cao, Feiming Bai, Naigang Wang, Jiefang Li, and D. Viehland
*Dept. of Materials Science and Engineering, Virginia Tech, Blacksburg, VA 24061*

Guangyong Xu and Gen Shirane
*Physics Department, Brookhaven National Laboratory, Upton, NY 11973*

Structural phase transformations of [110] electric field cooled Pb(Mg$_{1/3}$Nb$_{2/3}$)O$_3$-30%PbTiO$_3$ (PMN-30%PT) crystals have been performed by x-ray diffraction in a field-cooled (FC) condition. A phase sequence of cubic (C) → tetragonal (T) → orthorhombic (O) → monoclinic ($M_B$) was found on field-cooling (FC); and a R→$M_B$→O one was observed with increasing field beginning from the zero field-cooled (ZFC) condition at room temperature. The application of the [110] electric field induced a dramatic change in the phase sequence in the FC condition, compared to the corresponding data for PMN-30%PT crystals in a [001] field, which shows that the phase sequence in the FC condition is altered by the crystallographic direction along which a modest electric field (E) is applied. Only when E is applied along [110] are intermediate O and $M_B$ phases observed.

PACS number (s): 61.10Nz, 77.84.Dy, 77.80.Bh

## I. Introduction

Relaxor ferroelectric based morphotropic phase boundary (MPB) crystals, such as (1-x)Pb(Mg$_{1/3}$Nb$_{2/3}$O$_3$)-xPbTiO$_3$ (PMN-x%PT) and (1-x)Pb(Mg$_{1/3}$Nb$_{2/3}$O$_3$)-xPbTiO$_3$ (PMN-x%PT)[1], have attracted much interests as high performance piezoelectric actuator and transducer materials. For example, (001)-oriented PMN-33%PT crystals, which lie at the MPB, have the highest piezoelectric ($d_{33}$~2500 pC/N) and electromechanical coupling ($k_{33}$~94%)[2] coefficients. Following conventional thoughts[3], the MPB is supposed to be a vertical boundary between ferroelectric rhombohedral (R) and tetragonal (T) phases.

Park and Shrout[1,4] conjectured that the high electromechanical properties of PMN-x%PT and PZN-x%PT was due to a R→T phase transition induced by an applied electric field (E). More recently, an important breakthrough in understanding the structural origin of the high electromechanical properties of MPB compositions has been made. This is the discovery of ferroelectric monoclinic (M) phases bridging the R and T ones, which was first reported for Pb(Zr$_x$Ti$_{1-x}$)O$_3$[5-7]. Subsequently, x-ray (XRD) and neutron diffraction experiments have shown the existence of various ferroelectric M phases in oriented PZN-x%PT[8-12] and PMN-x%PT[12-15] crystals, including $M_A$ and $M_C$. Recent neutron diffraction studies of the effect of an applied E along [001] on the phase stability of PZN-8%PT by Ohwada *et al.*[11] have shown a R→$M_A$→$M_C$→T phase sequence with increasing E at 350K beginning from the zero-field-cooled (ZFC) condition, and a C→T→$M_C$ sequence in the field-cooled (FC) condition. Similar $M_A$ and $M_C$ phases have also been reported in PMN-x%PT[12-15]. A recent study by Bai *et al.*[15] established that PMN-30%PT has a C→T→$M_C$→$M_A$ sequence in the FC condition with the application of an electric field along the [001] direction, and a R→$M_A$→$M_C$→T one with increasing E beginning from the ZFC. The findings of prior diffraction studies of phase stability in [001] electric field cooled PMN-30%PT crystals are summarized in Figure 1(a).

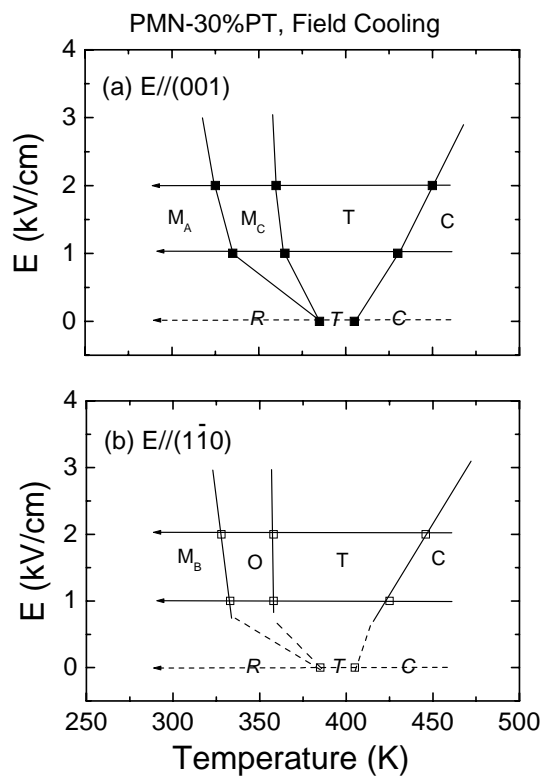

Figure 1. *E-T* phase diagram. Top panel (a) is PMN-30%PT in FC condition by Bai[15] with E along [001]; and bottom panel (b) is PMN-30%PT in FC condition with E along [110]. Arrows indicate the sequence of phase transition in FC condition. Dotted lines mean the ZFC condition, and solid lines mean FC condition.

The monoclinic symmetry allows the polarization vector to be unconstrained within a plane, rather than constricted to a particular crystallographic axis as for the higher symmetry R, T, or orthorhombic (O) phases. According to the polarization rotation theory[16], the high electromechanical properties of ferroelectric monoclinic phases are due to the rotation of the polarization vector within the symmetry allowed plane. Vanderbilt and Cohen predicted the stability ranges of monoclinic $M_A$ and $M_C$ phases using a thermodynamic approach [17]; in addition, they predicted a possible narrow stability range of a $M_B$ phase, intermediate between R and $M_C$ ones. Prior structural studies of PMN-x%PT and PZN-x%PT have only been performed under an E applied along the [001] direction – however, this is not an inherent restriction, as polarization rotation could occur in either direction in the permissible planes. For $BaTiO_3$, structural studies have been performed by Wada et. al.[18] under an E applied along the [111] direction, where a T→O→R phase sequence was observed with increasing E at 300K; and where optical birefringence indicated the presence of bridging M phases. Dielectric property studies of PMN-33%PT crystals with E along [110] have been reported by Lu et. al.[19], who reported an intermittently-present mestastable phase over a narrow temperature range sandwiched between $M_C$ and $M_A$ ones in the FC condition. Polarized light microscopy (PLM) indicated that this evasive phase was a single domain orthorhombic one [20]. In addition, the P-E and ε-E behaviors of ZFC PMN-30%PT crystals with E along [110] have been reported by Viehland and Li [21], whom conjectured a field-induced O phase at room temperature, via a monoclinic $M_B$ one. However, structural studies have not yet established this to be the case, nor has the phase sequence in [110] FC crystals yet been identified.

In this investigation, we have focused on establishing the structural transformation sequence of PMN-30%PT crystals with E along [110], and on determining how this sequence compares to that of the corresponding [001] FC one. Our results suggest that the phase sequences in a FC process with the electric field along different crystallographic orientations are distinctively different, as shown in Figure 1. XRD studies on PMN-30%PT under an [110] electric field have unambiguously shown a phase sequence of C→T→O→$M_B$ for [110] FC PMN-30%PT, and a sequence of R→$M_B$→O with increasing field beginning from the ZFC condition at room temperature.

## II. Experimental Procedure

Single crystals of PMN-30%PT with dimension of 3×3×3 mm$^3$ were obtained from HC Materials (Urbana, IL), and were grown by a top-seeded modified Bridgman method. The cubes were oriented along the pseudocubic ($1\bar{1}1$) / (110) / ($\bar{1}12$) planes, and were polished to 0.25μm. Gold electrodes were deposited on one pair of opposite (110) faces of the cube by sputtering. Temperature dependent dielectric constant measurements were performed using a multi-frequency LCR meter (HP 4284A) under various E.

The XRD studies were performed using a Philips MPD high-resolution system equipped with a two bounce hybrid monochromator, an open 3-circle Eulerian cradle, and a doomed hot-stage. A Ge (220)-cut crystal was used as an analyzer, which had an angular resolution of 0.0068˚. The x-ray wavelength was that of Cu$_{K\alpha}$=1.5406Å, and the x-ray generator operated at 45kV and 40mA. The penetration depth in the samples was on the order of 10 microns. The domain structure for PMN-30%PT under a [110] electric field can become quite complicated. In our diffraction studies, we have performed mesh scans around the (002) Bragg reflection in the (HHL) zone, defined by the [110] and [001] vectors; the (220) and ($2\bar{2}0$) reflections in the scattering zone defined by the [110] and [$1\bar{1}0$] vectors; and (200) in the (H0L) zone, defined by the [100] and [001] vectors. Each measurement cycle was begun by heating up to 550K to depole the crystal, and measurements subsequently taken on cooling. At 525K, the lattice constant of PMN-30%PT was a=4.027Å, correspondingly the reciprocal lattice unit (or 1 rlu) was a*=2π/a=1.560Å$^{-1}$. All mesh scans of PMN-30%PT with E along [110] shown in this study were plotted in reference to this reciprocal unit.

## III. Results
### III.1 Field-cooled condition
*Structural determination of different phases of various phase fields*

To determine the effect of E on the phase sequence, we measured changes in mesh scans on field-cooling under E=1, and 2kV/cm. At 450K under E=1kV/cm (data not shown), the (002) and (220) mesh scans did not exhibit splitting, and it was found that *c=a*. Thus it is clear that the lattice is cubic. As the temperature was decreased to 420K, the (002) reflection shifted towards slightly shorter wavevectors and a splitting along the longitudinal direction was found around the (200) reflection indicating a transition to the T phase. The signature of the T phase became more pronounced with decreasing temperature.

An electric field of 1kV/cm was then applied along the [110] direction. Figs. 2(a)-(d) show mesh scans taken around the (002), (200), ($2\bar{2}0$) and (220) for PMN-30%PT when the sample was cooled in the field to 375 K, respectively.

*Hu Cao e, al.*                                                    3

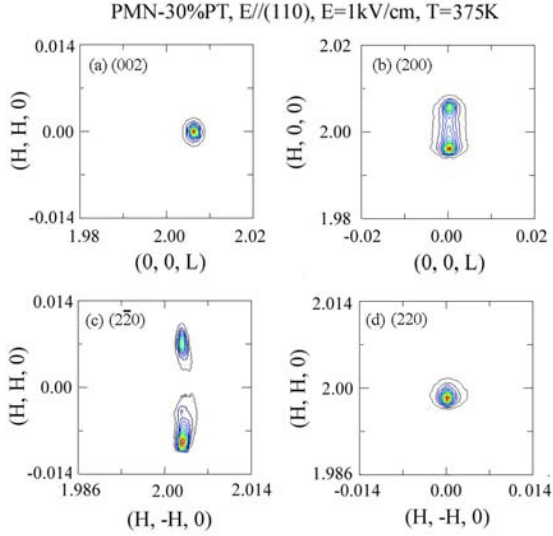
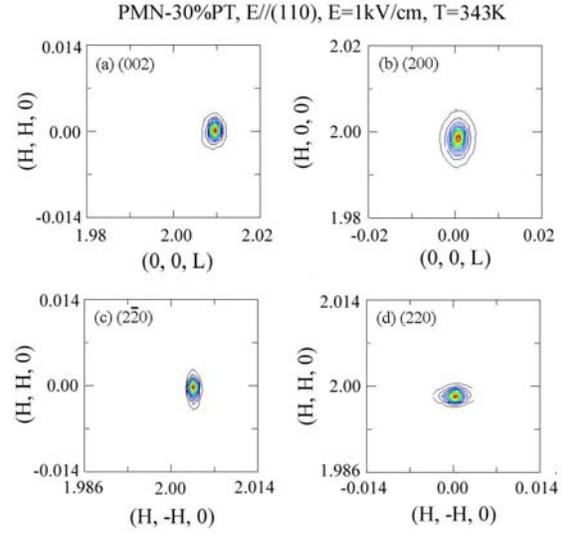

Figure 2. Mesh scans of (002), (200), $(2\bar{2}0)$, and (220) of PMN-30%PT with E=1kV/cm applied along [110] at 375K in FC condition.

Figure 3. Mesh scans of (002), (200), $(2\bar{2}0)$, and (220) of PMN-30%PT with E=1kV/cm applied along [110] at 325K in FC condition.

The (002) reflection (see Fig.2a) had only a single sharp peak. The lattice constant extracted from the (002) reflection was 4.0140Å. However, the (200) reflection (see Fig.2b) splits into two peaks along the longitudinal direction, from which the lattice parameters can be determined to be $a$=4.0142Å and $c$=4.0329Å, which are quite consistent with those found around the other zones in Figure 2. These results for PMN-30%PT with E along [110] reveal a tetragonal ferroelectric phase with twinned $a$- and $b$-domains along [200] or [020]. In addition, the (220) mesh scan (see Fig.2c) splits into two peaks only along the transverse direction, whereas the (220) scan (see Fig.2d) has only a single peak. This indicates that [110] field fixes the [110] crystallographic orientation; and that twinned $a$- and $b$- domains in the (001) plane are present in the $[2\bar{2}0]$ scan along the transverse direction. Thus, it is evident that PMN-30%PT has tetragonal lattice, in which the polarizations are constrained to [100] and [010] directions for $a$- and $b$-domains. As the temperature was decreased, the longitudinal splitting in the (200) mesh scan disappeared near 358K, indicating another phase transformation. Figures 3(a)-(d) show mesh scans taken about (002), (200), $(2\bar{2}0)$ and (220) within this phase field at 343K. Interestingly, only a single domain in each of these mesh scans was observed, indicating the presence of a well-developed single domain throughout the entire crystal. The structure of this phase was determined to be orthorhombic, where the polarization is fixed to the [110] direction. The lattice parameters of this orthorhombic phase were determined from the mesh scans to be $a_o$=5.6924Å, $b_o$=5.6812Å, and $c_o$=4.0070Å, where, $a_o$ was extracted from the (220) reflection, $b_o$ from the $(2\bar{2}0)$ one, and $c_o$ from the (002) one. This unit cell is a doubled one, consisting of two $M_C$ simple cells, as previously reported for orthorhombic $BaTiO_3$[22].

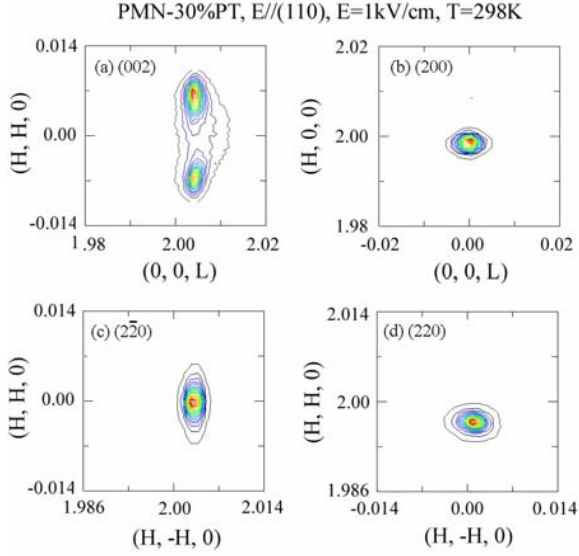

Figure 4. Mesh scans of (002), (200), $(2\bar{2}0)$, and (220) of PMN-30%PT with E=1kV/cm applied along [110] at 298K in FC condition.

Upon further decrease of the temperature to ~333K, the (002) mesh scan was found to split only along the transverse direction, revealing yet another phase transition. Figures 4(a)-(d) show (002), (200), $(2\bar{2}0)$, and (220) scans taken for E=1kV/cm at 298K. The (002) reflection (see Fig.4a) can be seen to split into two peaks with the same wavevector length; whereas, the other three mesh scans remained as a single peak. This is a signature of the monoclinic $M_A/M_B$ phase. We then determined the lattice parameters (by extraction from the mesh scans of Figure 4), to be $c_M$ =4.0204, $a_M/\sqrt{2}$ =4.0280Å, and $b_M/\sqrt{2}$ =4.0181Å; where $a_M$ and $b_M$ were derived from the (220) and $(2\bar{2}0)$ reflections, and $c_M$ from the (002) one. Our results show that on field-cooling below ~333K, a [110] field cooling can no longer sustain a single-domain O phase whose polarization is fixed to the [110] direction; rather, a transition to a polydomain monoclinic phase occurs. The unit cells of both the $M_A/M_B$ phases are doubled with respect to the primitive pseudocubic one, where the polarization lies in the $(1\bar{1}0)$ crystallographic plane. Although both the $M_A$ and $M_B$ phases belong to the *Cm* space group, there is different between their polarizations, for $M_A$, $P_x=P_y<P_z$; whereas for $M_B$, $P_x=P_y>P_z$. The fact $a_M/\sqrt{2}>c_M$ confirms that this monoclinic phase is the $M_B$ one. This is the first direct confirmation of the presence of the $M_B$ phase in the transformational sequence of either PMN-x%PT or PZN-x%PT single crystals. Although, it is relevant to note our prior reports of property data that indicate a R→$M_B$→O phase transformational sequence in the ZFC condition[21] and equally relevant to note that it is consistent with thermodynamic theory of Vanderbilt and Cohen [17] that allows for this sequence.

*Lattice parameters and dielectric behavior in FC condition.*

The lattice parameters of [110] electric field cooled PMN-30%PT at E=1kV/cm are plotted as a function of temperature in Fig. 5. The lattice parameter $a_C$ continuously decreases from 525K on cooling. At 428K, the value of $a_C$ began to gradually increase, indicating the formation of a small volume fraction of tetragonal phase. Near 415K, a splitting of the lattice parameter into $a_T$ and $c_T$ was observed and the crystal was completely transformed into the T phase, where $a_T$ was derived from the (002) reflection, whereas $c_T$ was derived from the (200) one. Here, one thing should be mentioned. The [110] field-cooled PMN-30%PT was a little different than the [001] field-cooled one, not only with regards to the domain configurations, but also with respect to the derivation of the lattice parameters. For example, in the T phase, $a_T$ was derived from the (200) reflection and $c_T$ from (002) reflection for [001] fielded PMN-30%PT. The C→T boundary shifted towards higher temperatures under a field of E=1kV/cm, relative to the ZFC condition. As the temperature was further decreased, subsequent phase transitions were observed. Between 358K and 333K, an orthorhombic phase was found with lattice parameters of $c_O$, $a_O$ and $b_O$. It is noteworthy that the values of $c_O$ and $a_T$, both determined from the (002) reflection, were continuous at the T→O transformation. The orthorhombic unit cell is a doubled one; and thus, the two lattice parameters, $a_O$ and $b_O$, are twice size of the corresponding monoclinic $M_C$ unit cell. The values of $a_O/\sqrt{2}$ and $b_O/\sqrt{2}$ shown in Fig. 5a exhibit a sharp decrease at the T→O transformation, relative to $c_T$. On further cooling below 333K, a transformation to a monoclinic $M_B$ phase was observed with three lattice $c_{M_B}$, $b_{M_B}/\sqrt{2}$, and $a_{M_B}/\sqrt{2}$. At the O→$M_B$ transformation, the values of $a_{M_B}/\sqrt{2}$ and $b_{M_B}/\sqrt{2}$ exhibited a sharp decrease, whereas $c_{M_B}$ had a sharp increase.

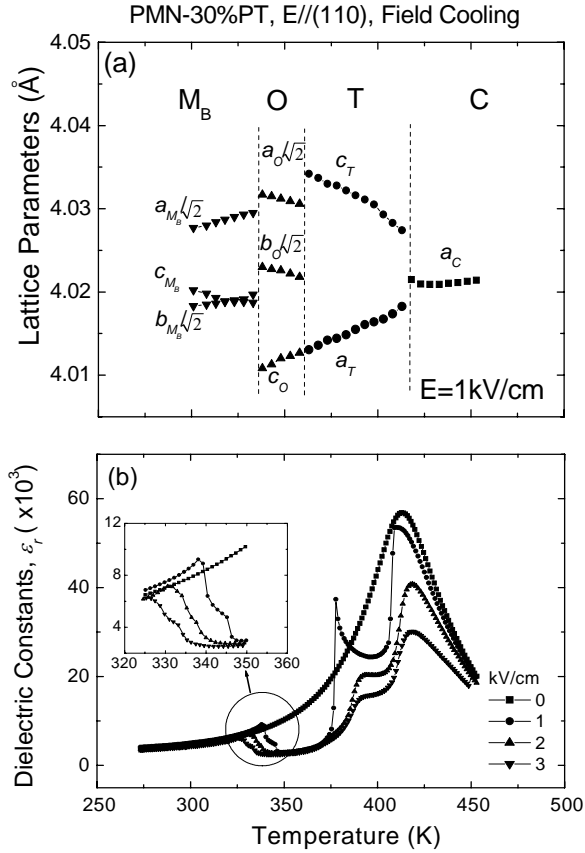

Figure 5. Temperature dependence of (a) lattice constants for PMN-30%PT with E=1kV/cm along [110], in which the lattice parameters $a_O/\sqrt{2}$, $b_O/\sqrt{2}$, and $c_O$; and $c_{M_B}$, $a_{M_B}/\sqrt{2}$ and $b_{M_B}/\sqrt{2}$ are plotted, and (b) dielectric behavior under various levels of electric field at $f$=1kHz in FC condition.

The dielectric behavior of [110] electric field cooled PMN-30%PT is shown in Fig. 5 (b). A plate-like sample (0.7mm thickness) was polished from the original cubic-shape crystal used in the XRD studies. The results are consistent with the transformational sequence of C→T→O→$M_B$ in the FC condition for various field levels. First, no significant shift of the value of the temperature of the dielectric maximum ($T_{max}$) was observed at the C→T transformation with increasing E. It is relevant to note that the C→T boundary as determined by $T_{max}$ did not shift with E, unlike that determined from the XRD data. We are currently investigating this difference, and it was not the purpose of the present work. The results of Fig.5b also show subsequent lower temperature phase transitions, corresponding to the T→O and O→M$_B$ ones observed by the lattice parameter changes in Fig. 5a. The value of the dielectric constant was relatively high in all phase fields, except in the single domain orthorhombic region, in which the polar vector is in coincidence with the direction of [110] applied E.

*Summary of domain-configurations in [110] electric field cooled T, O, and M$_B$ phases*

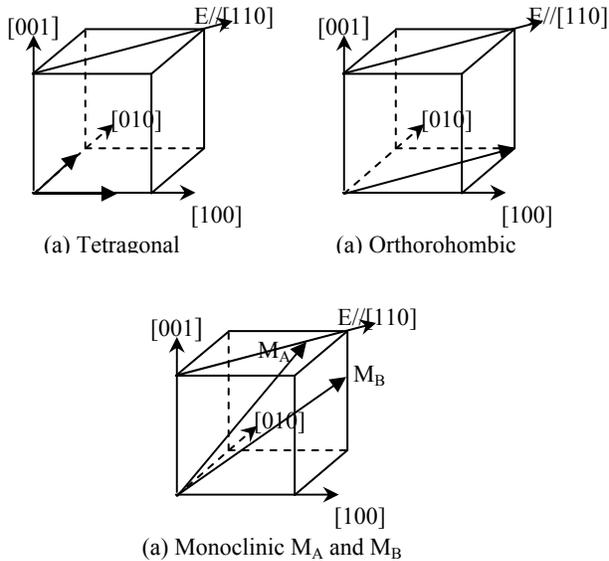

Figure 6. Schematic of tetragonal, orthorhombic and monoclinic M$_C$ phases in PMN-30% PT with E along [110]. Arrows indicate the polar vector. Panels (a) are twin domains developed in the tetragonal phase and only two polarizations are constrained within (001) plane; panel (b) shows the single domain developed in the orthorhombic phase; and panel (c) indicates the domains developed in the M$_B$/M$_A$ phase, in which the polarization vectors are constrained with the (110) plane.

Figure 6 conceptually summarizes the domain configurations of the T, O, and M$_B$ phases, in which an E has been applied along the [110] direction. In a ferroelectric T phase, six equivalent variants are permissible along the [100] direction. However, only two of these six are favored by applying an E along [110], which are [100] and [010]. In the T phase of [110] electric field-cooled PMN-30%PT, the [110] axis was fixed by the [110] field, and accordingly twinned *a*- and *b*-domains were found along [100] or [010] directions, as shown in Fig. 6a.

Figure 6c illustrates the single-domain orthorhombic state that is established throughout the crystal by [110] field cooling. Within this single domain, the polarization is fixed to only the [110] orthorhombic variant. There is no monoclinic tilting of this O variant away from the [110] direction. This unit cell is a doubled one, formed by perfectly-adjusting two M$_C$ simple cells.

The domain configurations of monoclinic phases are quite complicated; however, once an E is applied, a much simpler situation prevails[8]. For example, in the case of [001] electric field cooled PMN-30%PT [15] or PZN-8%PT [11], the field fixes the *c*-axis to lie along the pseudocubic [001] direction; thus, there are only two *b*-domains related by a 90° rotation around the *c* axis, each of which has two *a*-domains. These are M$_C$ (*a*-axis along [100]) and M$_A$ (*a*-axis along [110]) domain configurations previously reported for PMN-30%PT [15] and PZN-8%PT [11]. However, [110] field cooling may result in slightly more complicated domain configurations, as an E does not fix the *c*-axis to be along the [001] direction; rather, [110] field cooling fixes the [110] direction and forces the polarization as close as possible to the [110], as illustrated in Fig. 6c. However, in the monoclinic phase of [110] field cooled PMN-30%PT, the polarization is rotated away from the orthorhombic within the (1̄10) plane, pointing towards the [001].

This domain configurations is that of the monoclinic $M_B$ phase, since $P_x=P_y>P_z$ and $a_{M_B}/\sqrt{2} > c_{M_B}$. In this case, two polarizations were constrained to the $(1\bar{1}0)$ plane, consistent with a single peak in the (220) mesh scan and two domains in the (002) mesh scan.

## III. 2 Phase Stability with increasing E, beginning form ZFC

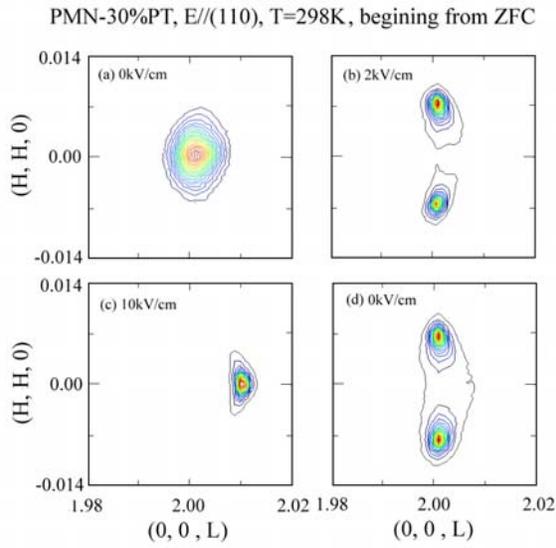

Figure 7.(002) mesh scans at 298K with increasing fields of (a) 0kV/cm, (b) 2kV/cm, (c) 10kV/cm, and (d) after removal of field in poled condition.

The field dependence of the lattice parameters was then investigated at room temperature, beginning from the ZFC condition. The crystal was first heated to 525K, and subsequently cooled under zero field. The (002) and (220) XRD mesh scans were obtained at various DC biases. Figures 7 (a)-(d) show the (002) scans for the field sequence of E=0kv/cm, 2kV/cm, 10kV/cm, and 0kV/cm (i.e., after removal of E) at 298K, respectively. For E=0kV/cm, only a single broad peak was found in the (002) scan, although a longitudinal splitting was observed in (220) scan (data not shown). The results show that the R phase is stable in the ZFC condition, with a lattice parameter of $a_R$=4.0220Å. Upon applied a field of 1kV/cm, a peak splitting was found to develop along the transverse direction in the (002) reflection, whereas the (220) scan only possessed a single peak (data not shown). These features are signatures of the monoclinic $M_B/M_A$ phase. The lattice parameters, $c_{M_B}$ and $a_{M_B}$, extracted from (002) and (220) reflections show that $a_{M_B}/\sqrt{2} > c_{M_B}$. Thus, we can conclude that the phase transformational sequence beginning from ZFC condition is R→$M_B$→O with increasing E, with the R→$M_B$ transition at E<1kV/cm and the $M_B$→O one near E=10kV/cm.

Figure 8 shows the electric field dependence of the lattice parameters at 298K. With increasing E to 8kV/cm, the value of $a_{M_B}/\sqrt{2}$ and $b_{M_B}/\sqrt{2}$ can be seen to continuously increase, exhibiting a sharp increment at 9kV/cm; whereas, the value of $c_{M_B}$ shows a gradual decrease and has a sharp decrease at 9kV/cm, at which point a single domain O phase is induced. Comparisons with the FC results in Figure 5a indicate that the crystal undergoes an abrupt transition to the O phase near 9kV/cm. It is also important to compare these results to recent studies of PMN-30%PT with E along [110] crystals by Li and Viehland[21], which indicated an induced phase transformation, near this same field in the ZFC condition. It is relevant to note that an hysteric P-E behavior was observed, whose remnant polarization was ~0.24Cc/m² (or $\frac{P_s}{\sqrt{3}}$) and whose value at the induced transition was ~0.3C/m² (or $\frac{P_s}{\sqrt{2}}$). It appears that the polarization can near continuously rotate within the (110) plane of the $M_B$ phase, from near the [111] direction to being in coincidence with the [110].

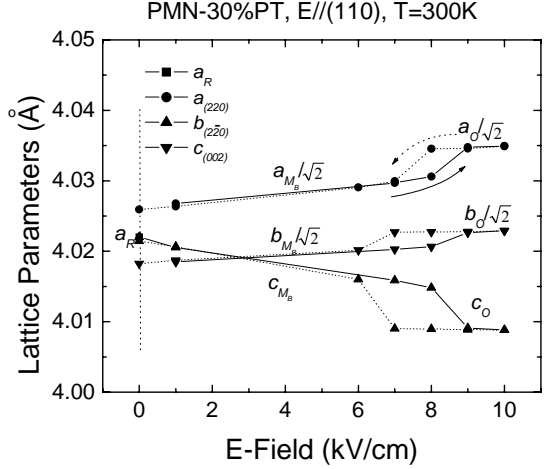

Figure 8. Electric-field dependence of the lattice parameters at 298K beginning from the ZFC condition, where lattice parameters $a_{(220)}/\sqrt{2}$, $b_{(2\bar{2}0)}/\sqrt{2}$, and $c_{(002)}$ are plotted. Solid represent data obtained on field-increasing, whereas dotted lines field-decreasing. At E=0kV/cm, $c_{(002)}=a_R$; and once E is applied $a_{(220)}/\sqrt{2}$, $b_{(2\bar{2}0)}/\sqrt{2}$ and $c_{(002)}$ correspond to $a_{M_B}/\sqrt{2}$, $b_{M_B}/\sqrt{2}$, and $c_{M_B}$ in the $M_B$ phase, and to $a_O/\sqrt{2}$, $b_O/\sqrt{2}$, and $c_O$ in the O phase, respectively.

With decreasing electric field between 9kV/cm and 6kV/cm, the lattice parameters revealed hysteresis of the induced $M_B \rightarrow O$ transformation. For E<6kV/cm, the orthorhombic phase did not remain stable, but rather a $M_B$ phase was recovered. In addition, for E<6kV/cm, the lattice parameters were equivalent between field increasing and decreasing sweeps. Upon removal of E, only a single domain was observed in the (220) scan, although a splitting along [220] was found in the (002) scan. We determined the monoclinic lattice parameters after removal of E and found that $a_{M_B}/\sqrt{2} > c_{M_B}$. These results show that the $M_B$ phase is the ground state condition for poled (110) crystals.

## IV. Summary

We summarize our findings in an E-T diagram concerning the ferroelectric stability of [110] oriented PMN-30%PT crystals in a [110] field, as shown in Fig. 1(b). For comparisons, a corresponding E-T diagram for the [001] field was given in Fig. 1(a). The C→T boundaries in the FC condition were nearly identical for both field orientations. In addition, the C→T boundaries in the FC condition were also nearly identical for both orientations, shifting by nearly the same degree with increasing E over the range studied. Significant difference between the [110] and [001] fields only became unambiguous on field-cooling to lower temperatures. For an E applied along the [110] direction, the sequence was found to be C→T→O→$M_B$; whereas for an E applied along the [001], it was C→T→$M_C$→$M_A$. In Figure 1(b), an intermediate O phase exists only when E is applied along the [110] direction, which constrains the polarization within the pseudocubic ($1\bar{1}0$) plane and the T→O boundary can be seen to be quite vertical for E<3kV/cm. A transition occurred with decreasing temperature in the FC condition, where the O→$M_B$ boundary shifted towards lower temperature with increasing E at a rate of 5K.cm/kV. Beginning from the ZFC condition at room temperature, the phase transformational sequence of R→$M_B$→O was observed with increasing E. Upon removal of electric field the crystal shows $M_B$ as the ground state.

Our results clearly demonstrate (i) the presence of intermediate ferroelectric orthorhombic and monoclinic $M_B$ phases in PMN-x%PT (or PZN-x%PT) crystals for the first time; (ii) a phase sequence of C→T→O→$M_B$ for [110] field cooled PMN-30%PT, and a sequence of R→$M_B$→O with increasing E along [110], beginning from the ZFC condition at room temperature.

**Acknowledgements**

We would like to gratefully acknowledge financial support from the Office of Naval Research under grants N000140210340, N000140210126 and MURI N0000140110761; U.S. Department of Energy under Contract No. DE-AC02-98CH10886. We would like to thank HC Materias for providing the single crystals used in this study. In particular, we would also like to acknowledge the tremendous help that Gen Shirane had provided us for along time, his presence will be missed, but his inspirations will always be present in our laboratory.